\documentclass[a4paper,11pt]{article}
\pdfoutput=1 
\usepackage{jheppub} 
\usepackage{amssymb}
\usepackage{graphicx}%
\usepackage{amsmath}%
\usepackage{color}

\newcount\timecount
\newcount\hours \newcount\minutes  \newcount\temp \newcount\pmhours
\hours = \time
\divide\hours by 60
\temp = \hours
\multiply\temp by 60
\minutes = \time
\advance\minutes by -\temp
\def\hour{\the\hours}
\def\minute{\ifnum\minutes<10 0\the\minutes
\else\the\minutes\fi}
\def\clock{
\ifnum\hours=0 12:\minute\ AM
\else\ifnum\hours<12 \hour:\minute\ AM
\else\ifnum\hours=12 12:\minute\ PM
\else\ifnum\hours>12
\pmhours=\hours
\advance\pmhours by -12
\the\pmhours:\minute\ PM
\fi
\fi
\fi
\fi
}

\def\monthname{\relax\ifcase\month 0/\or January\or February\or
March\or April\or May\or June\or July\or August\or September\or
October\or November\or December\else\number\month/\fi}

\def\bold#1{\setbox0=\hbox{$#1$}     \kern-.025em\copy0\kern-\wd0
\kern.05em\copy0\kern-\wd0
\kern-.025em\raise.0433em\box0 }

\def\URLtilde{\lower0.2em\hbox{$\tilde{\phantom{a}}$}}
\def\mycomm#1{\hfill\break\strut\kern-3em{\color{red}\tt ====> #1\color{black}}\hfill\break}

\def\beq{\begin{equation}}
\def\eeq{\end{equation}}

\def\ga{\mathrel{\raise.3ex\hbox{$>$\kern-.75em\lower1ex\hbox{$\sim$}}}}
\def\la{\mathrel{\raise.3ex\hbox{$<$\kern-.75em\lower1ex\hbox{$\sim$}}}}
\def\gev{{\rm \, Ge\kern-0.125em V}}
\def\tev{{\rm \, Te\kern-0.125em V}}
\def\gyr{{\rm \, G\kern-0.125em yr}}

\def\gappeq{\mathrel{\rlap {\raise.5ex\hbox{$>$}}
{\lower.5ex\hbox{$\sim$}}}}
\def\lappeq{\mathrel{\rlap{\raise.5ex\hbox{$<$}}
{\lower.5ex\hbox{$\sim$}}}}
\def\Toprel#1\over#2{\mathrel{\mathop{#2}\limits^{#1}}}

\def\m12{m_{1\!/2}}

\def\bea{\begin{eqnarray}}
\def\eea{\end{eqnarray}}

\def\beqar{\begin{eqnarray}}
\def\eeqar{\end{eqnarray}}

\def\m{{\cal m}}

\newcommand{\lsim}{\mathrel{\hbox{\rlap{\lower.55ex
\hbox{$\sim$}} \kern-.3em \raise.4ex \hbox{$<$}}}}
\newcommand{\gsim}{\mathrel{\hbox{\rlap{\lower.55ex
\hbox{$\sim$}} \kern-.3em \raise.4ex \hbox{$>$}}}}

\title{\boldmath Generalized Skyrmions in QCD\\ and the Electroweak Sector}
\author{John~Ellis$^{1,2}$,} \author{ Marek~Karliner$^3$} \author{and Michal Praszalowicz$^4$}
\affiliation{$^1$Theoretical Particle Physics and Cosmology Group, Physics Department,\\ 
King's College London, London WC2R 2LS, UK,\\
$^2$TH Division, Physics Department, CERN, CH-1211 Geneva 23, Switzerland,\\
$^3$Raymond and Beverly Sackler School of Physics and Astronomy,
Tel Aviv University, Israel,\\
$^4$M. Smoluchowski Institute of Physics,
Jagiellonian University, \\4 Reymonta Str., 30-059 Krak{\'o}w, Poland}

\emailAdd{John.Ellis@cern.ch}
\emailAdd{marek@neutron.tau.ac.il}
\emailAdd{michal@if.uj.edu.pl}

\abstract{We discuss the stability and masses of topological solitons in QCD and
strongly-interacting models of electroweak symmetry breaking
with arbitrary combinations of two inequivalent Lagrangian terms of fourth order in
the field spatial derivatives. We find stable solitons for only a restricted range of the ratio of these
combinations, in agreement with previous results, and we calculate the corresponding
soliton masses. In QCD, the experimental constraints on the fourth-order terms
force the soliton to resemble the original Skyrmion solution. However, this is not
necessarily the case in strongly-interacting models of electroweak symmetry
breaking, in which a non-Skyrmion-like soliton is also possible. This possibility will be
constrained by future LHC measurements and dark matter experiments.
Current upper bounds on the electroweak soliton mass range between 18 and 59~TeV,
which would be reduced to 4.6 to 8.1~TeV with the likely sensitivity of LHC data to the
fourth-order electroweak Lagrangian parameters.}

\begin{document}
\rightline{KCL-PH-TH/2012-40, LCTS/2012-24, CERN-PH-TH/2012-248,}
\rightline{TAUP-2958/12}

\maketitle
\flushbottom

\section{Introduction}

It is a generic feature of strongly-interacting theories that their vacua are
not invariant under all their global symmetries, and that the resulting
effective low-energy Lagrangians possess soliton solutions~\cite{W,ANW}. The archetype is
QCD, and many other four-dimensional examples have been studied theoretically~\cite{SW}.
In the case of QCD, most studies have built upon the pioneering work of
Skyrme~\cite{Skyrme}, and Skyrmion phenomenology has had a number of phenomenological
successes~\cite{SkyrmeOK,Meissner:1987ge,Weigel}.
However, 
generalizations of the Skyrme model are possible even in
QCD, and \textit{a fortiori} in other strongly-interacting models that may
exhibit dynamics rather different from QCD.

Strongly-interacting models of electroweak symmetry breaking have long
attracted considerable attention~\cite{TC}, with renewed intensity now that LHC
experiments have discovered a Higgs-like particle~\cite{X}. However, although it has
some of the characteristic properties predicted for the Higgs boson in the
Standard Model, the jury is still out, and many alternative scenarios
described by low-energy effective chiral Lagrangians remain viable~\cite{PNGB}. For
example, there is active interest in the simplest possibility that the
recently-discovered particle might be the pseudo-dilaton of some
nearly-conformal strongly-interacting electroweak sector~\cite{Dilaton}. Discriminating
between scenarios for electroweak symmetry breaking is a phenomenological
priority, and the existence (or otherwise) of soliton solutions may be a
valuable diagnostic tool for this task~\cite{EK}.

The masses and other properties of solitons in models described by effective
low-energy chiral Lagrangians depend on the strengths of higher-order terms in
their derivative expansions, and specifically of the coefficients of the
fourth-order terms~\cite{ANW}. In the case of the minimal effective Lagrangian for SU(2)
$\times$ SU(2) $\to$ SU(2) that may be used to describe both QCD and
electroweak symmetry breaking, there are two such parameters, 
as discussed in Section \ref{ClassMass}. One of these was
considered originally by Skyrme~\cite{Skyrme}, and has been the basis for most subsequent
studies of Skyrmion phenomenology~\cite{SkyrmeOK}. However, \textit{a priori} the other term
could also be present, and there have been some studies of generalized soliton
solutions in the presence of this extra term~\cite{PR,FOT,LLVC,DGH,AAN}. 
{In generic strongly-interacting models there may be a relation between the fourth-order Lagrangian parameters and the electroweak S parameter, but exploring this goes beyond the scope of this paper.}

Truncating the
derivative expansion at fourth order is dictated primarily by practical reasons, namely that phenomenological
information is available only about the fourth-order electroweak coefficients. However, we recall that
the contributions of higher-derivative terms to solitonic masses are not parametrically
suppressed~\cite{Aitchison:1986aq}.  
{In the leading order of the $1/N_{\rm c}$ expansion}
the soliton mass can be written as
an expansion in powers of the soliton size $r_0$:
\begin{equation}
M\sim 4 \pi \bar{v} N_{\rm c} \left[  a_{2}\bar{v}r_{0}+a_{4}\frac{1}{\bar{v}r_{0}}+a_{6}
\frac{1}{(\bar{v}r_{0})^{3}%
}+\ldots\right]  \, ,
\end{equation}
{where $\bar{v}=v/\sqrt{N_{\rm c}}$ is the $N_{\rm c}$-independent energy scale of 
chiral symmetry breaking (corresponding to $F_{\pi}/\sqrt{N_{\rm c}}$ in QCD),
and the $a_n$ are generic coefficients of $n$'th-order terms in the derivative expansion
that are expected to be of order 1}.
If one minimizes just the first two terms of $M$ with respect to $r_{0}$, treating
the sixth-order term as a perturbation, its
contribution is linear in $a_{6}$. This might suggest that
for large $a_{6}$ (and this is the case in QCD, where a
sixth-order term is generated by $\omega$-meson exchange 
\cite{Meissner:1987ge,Adkins:1983nw,Mashaal:1985rg}) 
the sixth-order term would even dominate. However, if one minimizes
the whole expression for $M$ (up to a
given order) the situation changes dramatically.   For positive  $a_6$, the sixth-order 
term provides a `barrier' around $r_{0}=0$ (as do higher-order
terms). Consequently, the optimal $r_{0}$ is increased and, although the 
$ a_{2}\bar{v}r_{0}$ term increases, the 4-th 
and 6-th order terms get smaller. As a result the sensitivity of the soliton mass to the 6-th order term
saturates at large $a_6>0$. The soliton mass
with large positive  $a_6$ is obviously larger than without it, but ``only" by factor
$\sim 2$ \cite{Mashaal:1985rg}, so
higher-order terms need not have large effects.

One should also remember that keeping only one specific 
fourth-order term, namely the Skyrme term introduced in Section~\ref{ClassMass}, gives
results accurate to within $20 - 30$\% in the case of QCD, which would be sufficient
for our purposes here. Estimates using the Skyrme term alone actually overshoot the
experimental nucleon mass, and we find that a somewhat larger range of classical masses is obtained
when the non-Skyrme term is included. We infer that the truncated Lagrangian may well
be a useful guide to possible upper limits on electroweak soliton masses.

The classical approximation is then
followed by semiclassical quantization. Semiclassical rotations are suppressed in QCD 
in the $1/N_{\rm c}$ expansion~\cite{ANW}. So even if a given Lagrangian generates
higher-order time derivatives, which is the case for the solitons with the non-Skyrme term
present, they can be neglected within large  $N_{\rm c}$ approximation.
A similar justification is applicable in many strongly-interacting 
models of electroweak symmetry breaking, but needs exploration on a case-by-case
basis. Here we do not study issues beyond the classical approximation.
Nor do we consider the negative Casimir ${\cal O}(N_{\rm c}^{0})$ contribution to the soliton mass,
which is the hardest to calculate. One can find discussion and estimates of the Casimir
energy in the QCD Skyrme model in Refs.~\cite{Moussallam:1992ia,Holzwarth:1995bv}.

As we review later in the context of our analysis, the existence or absence of
stable soliton solutions hinges upon the ratio of the two fourth-order
coefficients, and there is a generic range of this ratio where no stable
solitons exist~\cite{PR,FOT,LLVC,DGH,AAN}. Equally, there is a generic range, including the original
Skyrmion as a special case, where solitons do exist and are stable, at
least against spherically-symmetric decay. In the case of QCD, large-$N_{c}$ arguments
favour values of the fourth-order parameters within this stable range, close
to the Skyrmion limit, and this possibility is also favoured by the available
phenomenological estimates of these parameters in the effective chiral
Lagrangian of QCD~\cite{EGPR,Pichfit}.

However, currently we know very little about the possible magnitudes of the
fourth-order coefficients in the electroweak case, and it is possible that
their ratio is quite unlike the original Skyrme model, quite possibly in the
range where no stable solitons exist. On the other hand, if the
strongly-interacting electroweak sector is based on a theory with underlying
constituents that bind to form `electroweak baryons', one would expect
solitons to exist and describe qualitatively their masses and other
properties, even if they are rather different from the baryons and Skyrmions
in QCD-like theories~\cite{EK}.

The purpose of this paper is to explore the non-Skyrmionic possibilities~\cite{PR,FOT,LLVC,DGH,AAN},
particularly in the electroweak case, discussing their existence, stability
and masses in the classical approximation. This analysis is an essential
ingredient in the exploration of the consistency of different strongly-interacting electroweak
models with experiment. To this end, we assess the
prospects for probing strongly-interacting electroweak models by confronting present~\cite{DGPR,EGM} and
future limits~\cite{EGM} on (measurements of) fourth-order interaction parameters with
limits on (measurements of) electroweak baryon masses.
We also note that stable solitons would be present at some
level in the Universe today as relics from the Big Bang and contribute to the dark matter~\cite{Nussinov}. One should check
whether calculations of their abundance are compatible with cosmological and
astrophysical estimates of the density of cold dark matter, and whether
estimates of the rate for their elastic scattering on nuclei~\cite{CEO} are compatible
with upper limits from direct searches for dark matter~\cite{XENON100}. 

The structure of this paper is as follows. In Section~\ref{ClassMass} we discuss
the mass and stability of an SU(2) soliton in the classical limit, including both
possible fourth-order derivative terms. These calculations are used in
Section~\ref{pheno} to set phenomenological bounds on the possible masses
of QCD and electroweak solitons, taking into account the present and
prospective experimental constraints on the fourth-order terms. Section~\ref{dark}
summarizes considerations concerning electroweak solitons as cold dark matter,
and Section~\ref{conx} contains some concluding remarks.

\section{Classical Mass and Stability of an SU(2) Soliton}

\label{ClassMass}

The effective chiral Lagrangian corresponding to some strongly-interacting sector
with sponta\-neously-broken SU(2)$\times$SU(2) chiral symmetry is usually organized in terms
of (even) powers of derivatives of the chiral field:
\begin{equation}
U(x)=\exp\left(  i\frac{\vec{\tau}\cdot\vec{\pi}(x)}{v}\right) \, ,
\end{equation}
where the fields $\vec{\pi}(x)$ correspond to the Nambu-Goldstone bosons (pions in
the case of QCD, longitudinal polarization states of massive gauge bosons in the
case of a strongly-interacting electroweak sector) and $v$ is a typical symmetry breaking scale ($v=F_{\pi}=93$
MeV in QCD, $v\sim246$ GeV in the case of electroweak theory). The derivative
expansion is usually truncated at the fourth order, which could be reliable at energies below the
characteristic strong-interaction scale, $\sim$~GeV in QCD and $\sim$~TeV or more in the
electroweak theory:
\begin{equation}
\mathcal{L}_{\text{eff}}=\mathcal{L}_{2}+\mathcal{L}_{4} \, , \label{Leff}%
\end{equation}
where
\begin{equation}
\mathcal{L}_{2}=\frac{v^{2}}{4}\mathrm{Tr}\left(  \partial_{\mu}U\partial
^{\mu}U^{\dagger}\right)  . \label{Lkin}%
\end{equation}
In the SU(2) case there are two independent invariants containing just four
space-time derivatives~\cite{PR,FOT,LLVC,DGH,AAN}:
\begin{equation}
\mathcal{L}_{4}=2s\operatorname*{Tr}\left[  (R_{\mu}R_{\nu})(R^{\mu}R^{\nu
})-(R_{\mu}R^{\mu})^{2}\right]  +2t\operatorname*{Tr}\left[  (R_{\mu}R_{\nu
})(R^{\mu}R^{\nu})+(R_{\mu}R^{\mu})^{2}\right]  \, , \label{L4}%
\end{equation}
where%
\begin{equation}
R_{\mu}=\partial_{\mu}U\,U^{\dagger} \, .
\end{equation}
The parameters $s$ and $t$ can be in principle calculated from the underlying
strongly-interacting theory by integrating out the constituent degrees of freedom
(quarks and gluons in the case of QCD, and yet to be determined in an electroweak theory) or -- in a
phenomenological approach -- can be extracted from the data on the scattering
of the Nambu-Goldstone bosons or massive gauge bosons~\cite{EGM}~\footnote{For
rigorous lower bounds from Lorentz invariance, analyticity, unitarity and crossing, see~\cite{DGPR}.}. 
In QCD, as we discuss
in more detail later, the large-$N_{\rm c}$ expansion and pion data indicate
that $|t| \ll |s|$, so that the effective Lagrangian
contains (in a first approximation) only the kinetic term (\ref{Lkin}) and the first term in (\ref{L4}), as
in Skyrme's original work. For this reason, we refer to this as the Skyrme term,
and refer to other term in (\ref{L4}), that with coefficient $t$, as the non-Skyrme term.

Over fifty years ago already, Skyrme~\cite{Skyrme} observed that (\ref{Leff}) with $t=0$
possesses solitonic solutions that, due to
the fact that $\pi_{3}($SU(2)$)=\mathbb{Z}$, carry an integer-valued topological quantum number 
\begin{equation}
B=\frac{1}{24\pi^{2}}\int d^{3}x\epsilon^{ijk}\mathrm{Tr}\left[  (U^{\dagger
}\partial_{i}U)(U^{\dagger}\partial_{j}U)(U^{\dagger}\partial_{k}U)\right]  .
\label{Bnumber}%
\end{equation}
that can be interpreted as baryon number (see also~\cite{W}).

It is straightforward to calculate the classical contribution to the 
mass of the baryon in terms of the parameters $v$, $s$ and $t$.
In principle, all terms in the derivative expansion make significant contributions to the soliton mass, 
whereas we have no information on the possible magnitudes of higher-order terms. 
On the other hand, we know that classical calculations in QCD keeping only the fourth-order term 
Skyrme term, i.e., setting $t = 0$, are accurate to $\sim 20 - 30$\%~\cite{SkyrmeOK,Weigel}. 
In the case with non-zero non-Skyrme term a somewhat larger masses are allowed,
cf. our calculation below for QCD point A in Sect \ref{phenoQCD}. 
We therefore hope that classical calculations in the electroweak 
case, keeping the general form of fourth-order coupling, sample 
the range of possible masses with similar accuracy.

\begin{figure}[h]
\centering
\includegraphics[scale=0.40]{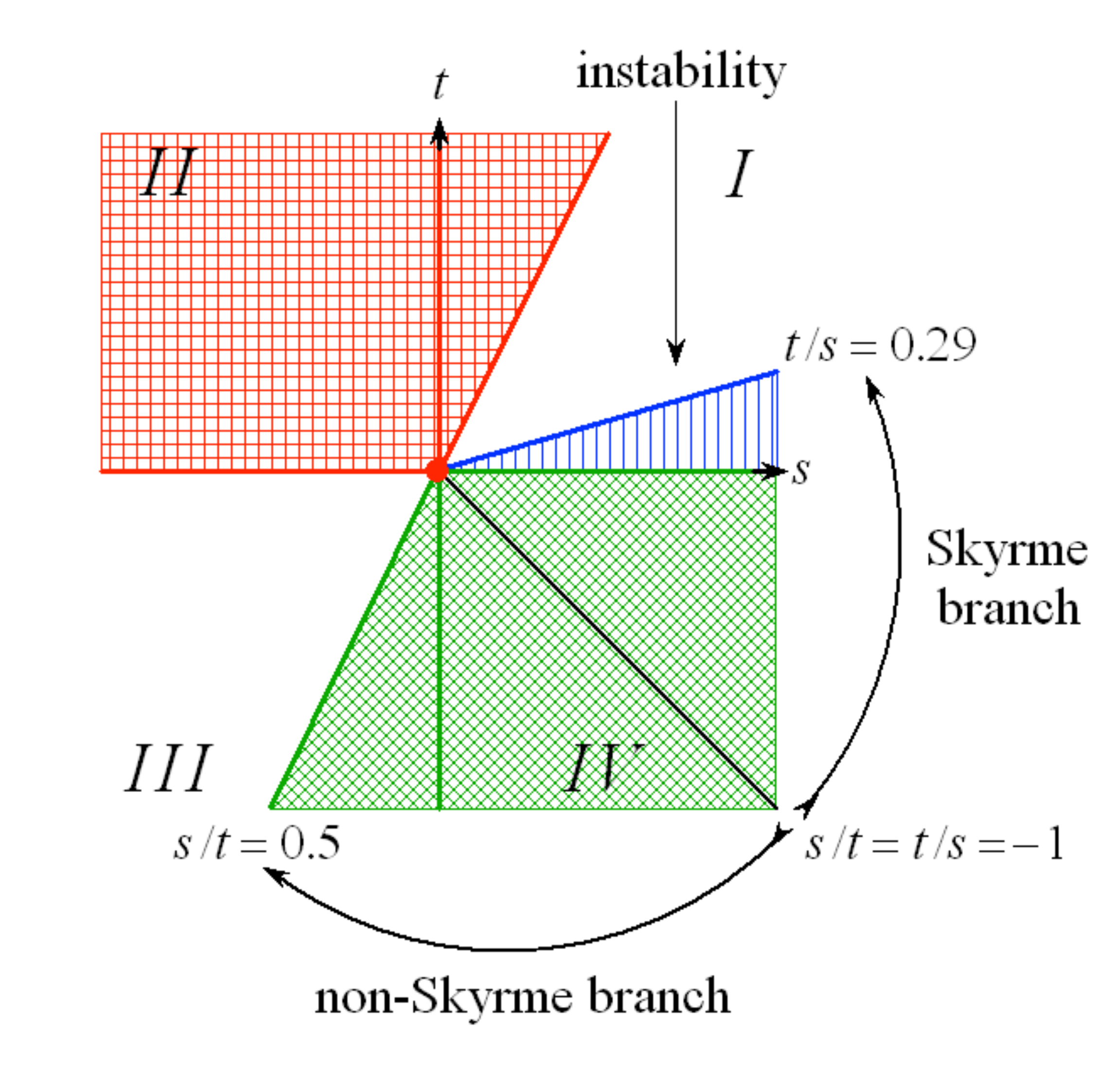} \quad\caption{\it 
The $(s,t)$ parameter plane, indicating with green shading (diagonal squares)
the range of 
parameters satisfying the positivity bound
$M_{4}>0$ and hence admitting a solution of the differential equation (\ref{rownanie}),
and with red shading (vertical squares)
the range where solutions do not exist due to the negativity of the
fourth order term: $M_{4}<0$. 
In the range $t/s<0.29$ within the first quadrant, 
indicated in blue (vertical bars), 
solutions with positive $M_{4}$ exist, but no positivity bound can be
derived.}%
\label{tsplane}%
\end{figure}

Possible refinements to this classical calculation could include calculation of the Casimir energy, 
inclusion of the current-quark mass terms and semiclassical quantization of the rotational modes
(which provide better agreement with the observed baryon masses). However, we do not
go into them here, for three reasons in addition to our ignorance of
possible higher-order terms in the electroweak case. One is that we are primarily
interested in bounds on the masses of the possible electroweak
baryons, rather than in details of their spectrum. A second reason is that,
in any case, we lack the information
about the underlying strongly-interacting theory that would be needed to calculate the
non-classical corrections in this case. The third reason is that in reality, even in QCD, 
the parameter $t$ is not necessarily negligible. For example in an approximation where
the Skyrme Lagrangian is derived by integrating out heavy mesons (typically $\rho$ and $\sigma$)
one can relate the parameters $s$ and $t$ to the heavy meson masses 
\cite{Meissner:1987ge,Mashaal:1985rg,Pham:1985cr}. Comparing our Lagrangian
(\ref{Leff}) with Ref.~\cite{Pham:1985cr} we get:
\begin{equation}
s=\frac{F_{\pi}^{2}}{192}\frac{3m_{\sigma}^{2}-2m_{\rho}^{2}}{m_{\rho}%
^{2}\,m_{\sigma}^{2}}=(0.6\div3.2)\times10^{-4},\qquad t=\frac{F_{\pi}^{2}%
}{96m_{\sigma}^{2}}=(5.8\div8.4)\times10^{-4} \label{strel}%
\end{equation}
with $F_{\pi}=186$~MeV.
The soliton is unstable with these values, as we discuss shortly in connection with Fig.~\ref{tsplane},
since $t/s>1.7$ and stable solutions exist only for $t/s \le 0.29$. Moreover, the values (\ref{strel})
are beyond the range allowed for the QCD effective Lagrangian discussed in Sect.~\ref{phenoQCD}.
Therefore, it does not seem possible to constrain the electroweak $s$ and $t$ parameters using
ideas about the (as yet unknown) technimeson
masses~\footnote{We note also that (\ref{strel}) implies $t>0$, whereas the
bounds shown in Fig.~\ref{figst_vs_alfas_QCD} do not exclude negative $t$.}.

Our primary interest here will be
the classical soliton mass in the presence of  a non-negligible non-Skyrme term, and
how the possibility that $t \ne 0$ affects the (approximate) mass bounds provided by the
Skyrme calculation with $t = 0$ discussed in~\cite{EK}.

Classical soliton solutions of the chiral field equations are usually found within the spherically-symmetric
`hedgehog' Ansatz for a static field configuration:
\begin{equation}
U(\vec{r}\,)=\exp\left(  i\frac{\vec{\tau}\cdot\vec{r}}{r}P(r)\right) \, ,
\label{hedgehog}%
\end{equation}
where the profile function $P(r)$ is a solution to the Euler-Lagrange equation
of motion, which for the Lagrangian (\ref{Leff})--(\ref{L4}) takes the following form:%
\begin{equation}
P^{\prime\prime}(\rho)=-\frac{G(\rho)}{F(\rho)} \label{rownanie}%
\end{equation}
with boundary conditions%
\begin{equation}
P(0)=\pi,\;P(\infty)=0 \, .
\end{equation}
Here%
\begin{align}
F(\rho)  &  \equiv \; \rho^{2}\left(  1-96\,tP^{\prime\,2}\right)  +64\,s\sin
^{2}P \, ,\nonumber\\
G(\rho)  &  \equiv \; -64\,t\rho P^{\prime\,3}+32\,s\sin2P\,P^{\prime\,2}+2\rho
P^{\prime}\nonumber\\
&  \; -\sin2P\left(  1+32\left(  s-t\right)  \frac{\sin^{2}P}{\rho^{2}}\right) \, ,
\label{FandG}%
\end{align}
and the dimensionless radial variable $\rho$ is defined as%
\begin{equation}
\rho\; \equiv \; rv \, ,
\end{equation}
where $v = f_\pi, \simeq 246$~GeV in the QCD and electroweak cases, respectively.

Following the pioneering paper~\cite{ANW}, classical
solutions of the field equations were studied numerically long ago in~\cite{PR}
and stability regions in the $(s,t)$ parameter space were analyzed
in~\cite{FOT,LLVC,DGH,AAN}.
We have repeated their analyses for the purpose of the present work with
the following results. 

The total classical soliton mass%
\begin{equation}
M_{\text{sol}}=4\pi v\left(  M_{2}+M_{4}\right)
\end{equation}
where
\begin{align}
M_{2}  &  =\frac{1}{2}%
{\displaystyle\int\limits_{0}^{\infty}}
d\rho\left(  \rho^{2}P^{^{\prime}\,2}+2\sin^{2}P\right) \, , \nonumber\\
M_{4}  &  =16s%
{\displaystyle\int\limits_{0}^{\infty}}
d\rho\left(  2P^{\prime\,2}\sin^{2}P+\frac{\sin^{4}P}{\rho^{2}}\right)  -8t%
{\displaystyle\int\limits_{0}^{\infty}}
dr\left(  \rho^{2}P^{\prime\,4}+2\frac{\sin^{4}P}{\rho^{2}}\right) \, ,
\label{mtot}%
\end{align}
where the contribution $M_{2}$ corresponds to $\mathcal{L}_{2}$
and is clearly always positive,
and the contribution $M_{4}$ corresponds to $\mathcal{L}_{4}$.
By changing variable once more: $\rho\rightarrow
\rho/\sqrt{\left\vert s\right\vert }$ or $\rho\rightarrow\rho/\sqrt{\left\vert
t\right\vert }$, one can show that the solutions depend only on the ratio
$t/s$. By rescaling: $\rho\rightarrow\lambda\rho$, one can show that
solutions exist only when $M_{4}$ is positive.
Inspecting directly $M_{4}$ in (\ref{mtot}), we see that in the fourth quadrant
of the $(s,t)$ plane ($s>0,$ $t<0$) $M_{4}$ is always positive, so that solutions
of (\ref{rownanie}) always exist, whereas in the second quadrant ($s<0,$
$t>0$) $M_{4}$ is always negative, so there are no solutions
of (\ref{rownanie}). In the first and third quadrants
positivity bounds can be derived~\cite{FOT,PR}. It turns out that in the first
quadrant $M_{4}$ is negative for $t/s>2$, so the soliton is unstable, and in the third quadrant $M_{4}$ stays
positive for $t/s>2$, so the soliton is stable. This is illustrated in Fig.~\ref{tsplane}, where the regions
of positive and negative $M_{4}$ are displayed by green shading (diagonal squares)
and red shading (vertical squares), respectively.

In the remaining parts of quadrants $I$ and $III$, no positivity bounds can be
derived and the positivity of $M_{4}$ has to be checked numerically. It turns
out that in the third quadrant $M_{\text{sol}} > 0$ and $\to 0^+$ when
$t/s\rightarrow2^+$, whereas no solution exists for $t/s<2$. The situation is
different in the first quadrant, where solutions with positive $M_{4}$
exist for small $t/s$ below the line $t/s \simeq 0.29$, where the function $F(\rho)$
vanishes and equation (\ref{rownanie}) cannot be solved. Interestingly,
$F(\rho)$ is equal to the second variation of $M_{\text{sol}}$, so the positivity of $F(\rho)$
is a necessary condition for $M_{\text{sol}}$ to be the minimum with respect to
the variations of $P(r)$. Below the line $t/s\sim0.29$ the solution is classically stable,
and above this line it is classically unstable.

However, it has been shown in~\cite{PR} that there is no lower limit on the
soliton mass anywhere in the first quadrant. This means that, when $t/s < 0.29$,
the classical solution discussed above, though stable against local
spherically-symmetric perturbations, can only be metastable at best, and might be
unstable against non-spherically-symmetric perturbations. This region is shaded
blue (with vertical bars) in Fig.~\ref{tsplane}.

\begin{figure}[h]
\centering
\includegraphics[scale=1.1]{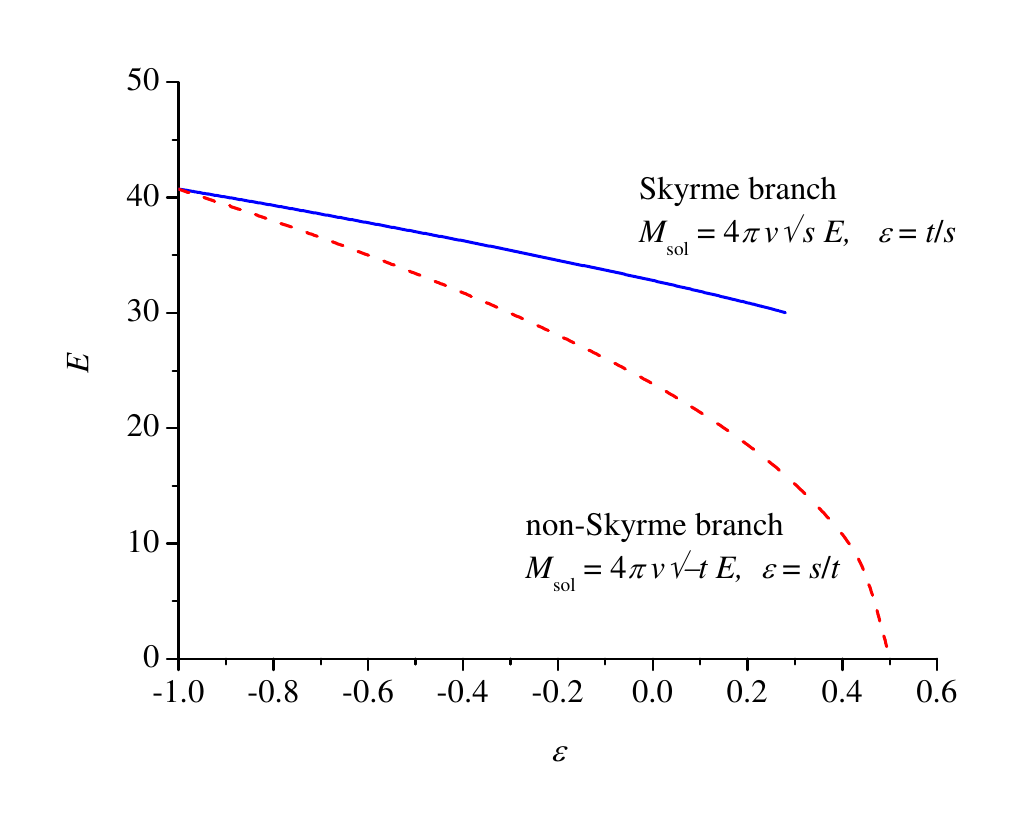} \quad\caption{\it 
The soliton mass 
in units
of $4\pi v \sqrt{s}$ for the Skyrme branch where 
$\epsilon=t/s$ (solid blue line),
and for the the non-Skyrme branch in units of $4\pi v \sqrt{-t}$ where
$\epsilon=s/t$ (dashed red line). The Skyrme branch ends at $t/s \sim 0.29$
(see Fig. \ref{tsplane}).}%
\label{figmass}%
\end{figure}

As already remarked, soliton solutions depend only on
the ratio $t/s$ (or $s/t$). We analyze solutions starting from
the Skyrme-like case with $t=0$
and the non-Skyrme-like case with $s=0$,
and have divided the allowed parameter space into two branches
on either side of the line $t/s = -1$, as
also depicted in Fig.~\ref{tsplane}. In order to solve equation (\ref{rownanie})
within the Skyrme branch where $s\neq0$, we have rescaled $\rho\rightarrow
\rho/\sqrt{s}$ and evaluated the soliton mass in units of $4\pi v\sqrt{s}$,
whereas within the non-Skyrme branch where $t\neq0$ we have rescaled
$\rho\rightarrow\rho/\sqrt{-t}$ and evaluated the soliton mass in units of
$4\pi v\sqrt{-t}$. The results are presented in Fig.~\ref{figmass} as
functions of $\varepsilon \equiv t/s$ for the Skyrme branch and $\varepsilon \equiv s/t$ for
the non-Skyrme branch.

\begin{figure}[h]
\centering
\includegraphics[scale=0.40]{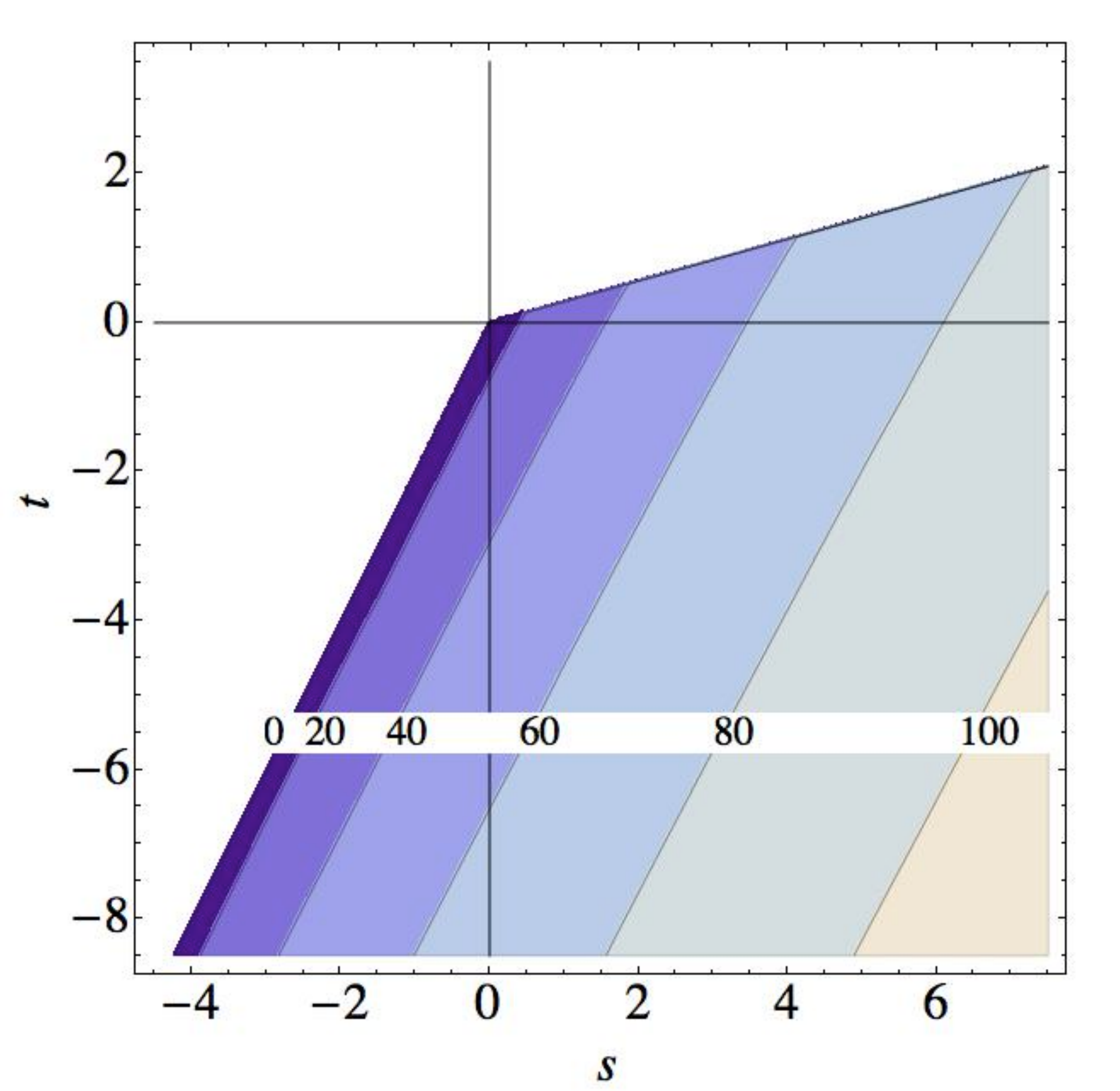} \quad\caption{\it Contour plot of the
mass of the generalized Skyrmion in units of $4 \pi v$.}%
\label{figlevels}%
\end{figure}

Finally, in Fig.~\ref{figlevels} we present a contour plot of the classical
soliton mass in units of $4\pi v$ over the allowed range of parameter space. We
see that the contours of constant $M_{\text{sol}}$ are almost parallel to the
line $t/s=2$ where the soliton mass vanishes. This feature is very helpful for finding the maximum of the
generalized soliton mass, as we discuss in the next Section. 

\section{Phenomenological Estimates of Soliton Masses}
\label{pheno}

We now discuss constraints on the conventional baryon mass
using constraints on the parameters of the low-energy effective chiral Lagrangian
for QCD, and constraints on the masses of possible `electroweak baryons' given by
the constraints on higher-order electroweak Lagrangian parameters
from electroweak theory, current phenomenology and
the potential sensitivity from the LHC~\cite{EGM,EK}. 

Phenomenological constraints
on higher-order Lagrangian parameters are often given in terms of the
coefficients $\alpha_{4,5}$ that are related to the 
parameters $s$ and $t$ discussed in Section~\ref{ClassMass} in the
following way \cite{EK}:%
\begin{equation}
s \; = \; \frac{\alpha_{4}-\alpha_{5}}{4},\;t \; = \; \frac{\alpha_{4}+\alpha_{5}}{4} \, .
\end{equation}
Although limits on electroweak baryon masses are our principal interest, we first discuss the limits
that can be derived for ordinary baryons in QCD, comparing the non-Skyrmion case
with $t \ne 0$ with the conventional Skyrmion case $t = 0$.

\subsection{QCD baryons}
\label{phenoQCD}

\begin{figure}[h]
\centering\includegraphics[scale=0.45]{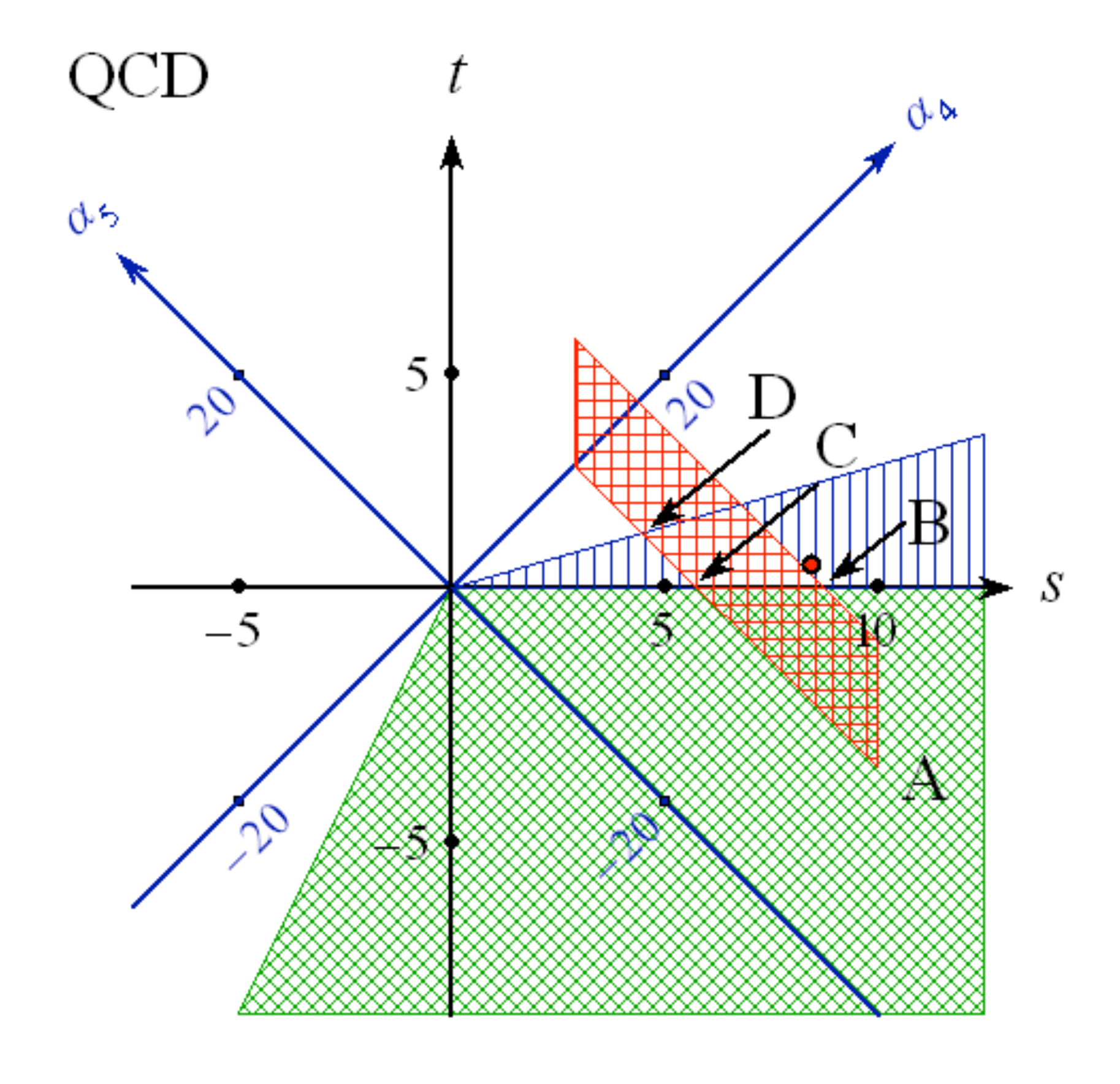} \quad
\caption{\it Comparison of  the parameter ranges allowed for the low-energy
chiral perturbation theory coefficients in QCD~\cite{EGPR,Pichfit},
eq.~(\ref{QCDbounds}),
with the range of parameters where stable baryonic solitons
exist.
The parameters $\alpha_{4}$, $\alpha_{5}$, $s$ and $t$ are in units of
$10^{-4}$. The red spot corresponds to the large-$N_{\rm c}$ values 
given in eq.~(\ref{LargeNc}).
Point A has the maximal mass overall, Points B and C bracket the range of 
parameters allowed in the pure Skyrme limit: $t = 0$, and Point D lies on the stability
boundary $t/s = 0.29$ and has the minimal mass overall.
}%
\label{figst_vs_alfas_QCD}%
\end{figure}

Higher-order coefficients in the low-energy effective Lagrangian
of QCD have been the subject of many studies. Here we
use the following ranges extracted from low-energy strong-interaction data:
\cite{EGPR,Pichfit,EK}:%
\begin{align}
11\times10^{-4} &  <\,\,\,\,\,\,\,\alpha_{4}\,\,\,\,\,\,<17\times10^{-4} 
\, ,\nonumber\\
14\times10^{-4} &  <\alpha_{4}-\alpha_{5}<40\times10^{-4}.\label{QCDbounds}%
\end{align}
We can compare these values with the predictions of the
large-$N_{\rm c}$ approximation within the framework of chiral SU(3) $\times$ SU(3) $\to$ SU(3)~\cite{Pichfit}:
\begin{eqnarray}
\alpha_{4} & = & \phantom{{-}}18 \times 10^{-4} \, ,\nonumber\\
\alpha_{5} & = & {-}16 \times10^{-4} \, .\label{LargeNc}%
\end{eqnarray}
The bounds (\ref{QCDbounds}) are displayed in Fig.~\ref{figst_vs_alfas_QCD},
where the large-$N_{\rm c}$ prediction is also indicated, as a red spot~\footnote{We represent the constraints
(\ref{QCDbounds}), (\ref{currentEWrange}), (\ref{possibleLHCbounds}) as parallelograms in
the $(s, t)$ plane, whereas they should be ellipses. However, the correlations between the
errors in $s$ and $t$ are not available, so these ellipses cannot be drawn accurately.}.

As already discussed in Section~\ref{ClassMass} the contours of constant
$M_{\text{sol}}$ are almost parallel to the line $t=2s$. Therefore the maximal
mass corresponds to the right-most corner of the region allowed by the QCD bounds,
called Point A below. For
completeness, we also include two pure Skyrme points (B and C) and the 
point corresponding to minimal mass, located right below the stability line $t/s = 0.29$ (Point D):

\begin{itemize}
\item $\mathrm{A}: (s, t) = (10, - 4.5)\times10^{-4}$ (maximal mass point
overall):
$M_{\mathrm{A}}\simeq1354$ MeV;

\item $\mathrm{B}: (s, t) = (8.5, 0.0)\times 10^{-4}$ (maximal mass Skyrme point):
$M_{\mathrm{B}}\simeq1118\;$MeV;

\item $\mathrm{C}: (s, t) = (5.5, 0.0)\times 10^{-4}$ (minimal mass Skyrme point):
$M_{\mathrm{C}}\simeq900\;$MeV$.$

\item $\mathrm{D}: (s, t) = (4.3, 1.2)\times 10^{-4}$ (minimal mass point
overall):
$M_{\mathrm{D}}\simeq728\;$MeV$.$
\end{itemize}

We see that the range of classical masses allowed in the Skyrme case $t = 0$
includes the physical value of the nucleon mass. In terms of the conventional
representation
\begin{equation}
s \; = \; \frac{1}{32e^{2}} \, ,%
\end{equation}
this range corresponds to
\begin{equation}
6 \; < \; e \; < \; 7.5 \, .
\end{equation}
We also note that the range of possible QCD baryon masses is extended
significantly in the presence of a non-Skyrme term, by ${\cal O}(200)$~GeV
in either direction. However, detailed QCD baryon phenomenology including
the evaluation of  corrections to the classical soliton mass due to the Casimir energy, 
inclusion of the current-quark mass terms and semiclassical quantization of the rotational modes
lies beyond the scope of this paper.

\subsection{Current bounds on electroweak baryon masses}

\begin{figure}[h]
\centering
\includegraphics[scale=0.35]{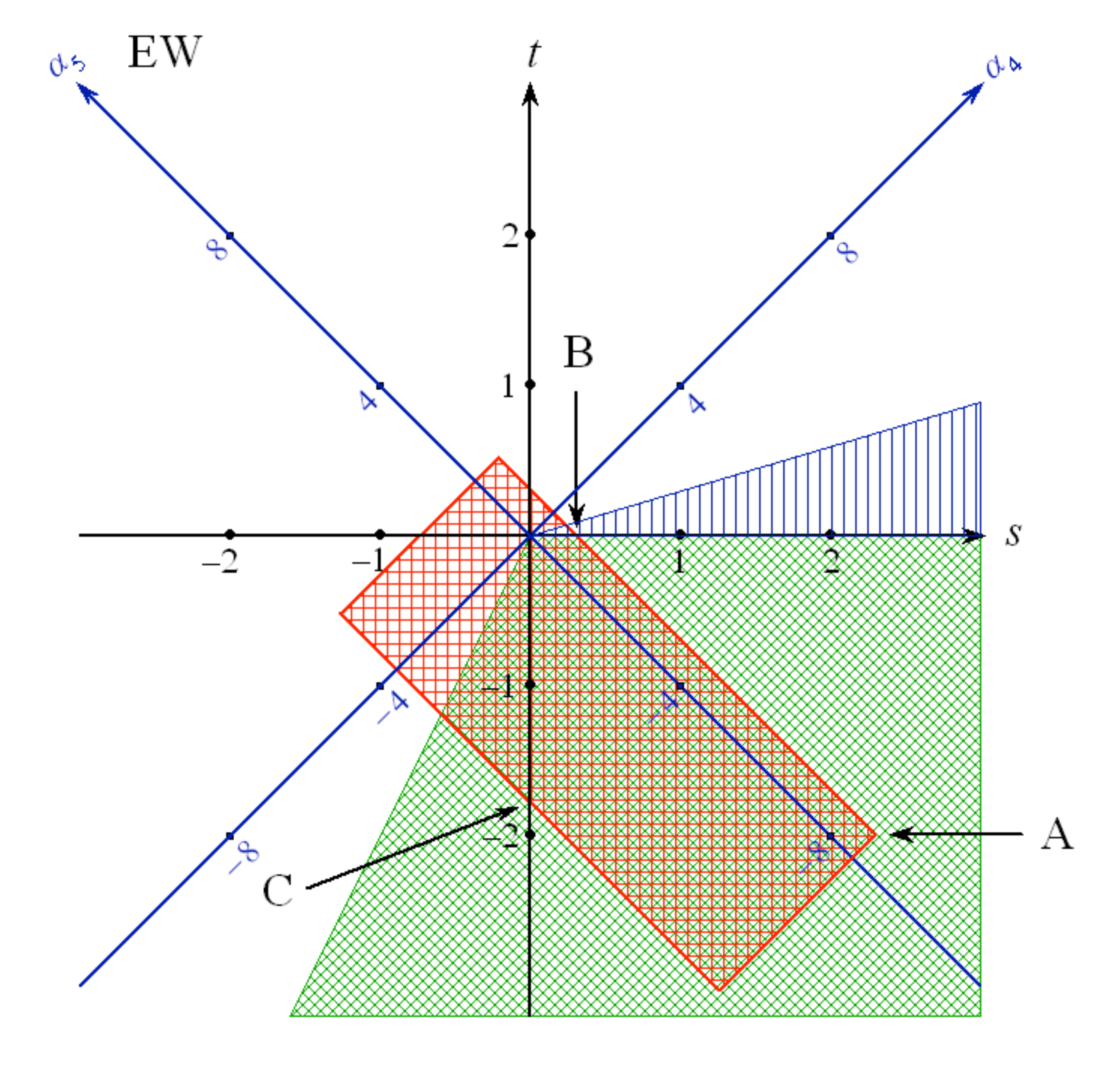} \quad\caption{\it Comparison 
of  the parameter ranges currently allowed~\cite{EGM} for the effective low-energy
electroweak Lagrangian with the range of parameters where stable electroweak baryonic solitons
exist. The parameters $\alpha_{4}$, $\alpha_{5}$, $s$ and $t$ are in units of
$10^{-1}$. Point A has the maximal mass overall, Point B has the
maximal mass 
allowed in the pure Skyrme limit: $t = 0$, and Point C has the
maximal mass allowed
in the limit where the Skyrme term vanishes: $s = 0$.}%
\label{figst_vs_alfas_EW}%
\end{figure}

The higher-order coefficients in an effective electroweak Lagrangian are currently
poorly constrained, namely by the following bounds \cite{EGM}:%
\begin{align}
-3.5\times10^{-1} &  <\alpha_{4}<0.6\times10^{-1} \, ,\nonumber\\
-8.7\times10^{-1} &  <\alpha_{5}<1.5\times10^{-1} \, . \label{currentEWrange}%
\end{align}
These constraints are superimposed on the $(s,t)$ plane in Fig.~\ref{figst_vs_alfas_EW}.
As in the QCD case, we superpose various points that illustrate the range of
possible electroweak baryon masses currently allowed, as obtained
using the values of the soliton mass
calculated in Section~\ref{ClassMass} and displayed in Fig.~\ref{figmass}. We note
that there is no minimal electroweak baryon mass, since the constraints
(\ref{currentEWrange}) include the possibility that $M_{\rm sol}=0$ for $t=2 s< 0$. 
The points displayed include
the maximal mass point denoted by A, the maximal-mass Skyrme case (Point B),
and the maximal-mass case with vanishing Skyrme term $s = 0$ (Point C):

\begin{itemize}
\item $\mathrm{A}: (s, t) = (0.23, {-}0.20)$ \ (maximal mass point
overall): \ $M_{\mathrm{A}%
}\simeq59$ TeV;

\item $\mathrm{B}: (s, t) = (0.03, \phantom{{-}}0.0\ )$ \ (maximal mass Skyrme point): $M_{\mathrm{B}}%
\simeq18\;$TeV;

\item $\mathrm{C}: (s, t) = (0.0,\,\, {-}0.175)$ (maximal mass with $s=0$):
\ \ \ $M_{\mathrm{C}%
}\simeq31\;$TeV$.$
\end{itemize}

This analysis indicates that the upper bound on the possible electroweak baryon mass over
the range of parameter space currently allowed is approximately $60$~TeV~\footnote{Assuming
that the model-dependent corrections to the classical mass calculation are not very large.}. This upper limit is
somewhat higher than in~\cite{EK}, because we have extended the analysis to the case of a non-zero
non-Skyrme term. 
If the conjectural strongly-interacting electroweak
theory were to resemble QCD, i.e., the parameter $t$ were small,
the upper limit on the possible electroweak baryon mass limit would be significantly
reduced to $\simeq 18$~TeV.

\subsection{Prospective LHC bounds on electroweak baryon masses}

\begin{figure}[h]
\centering\includegraphics[scale=0.35]{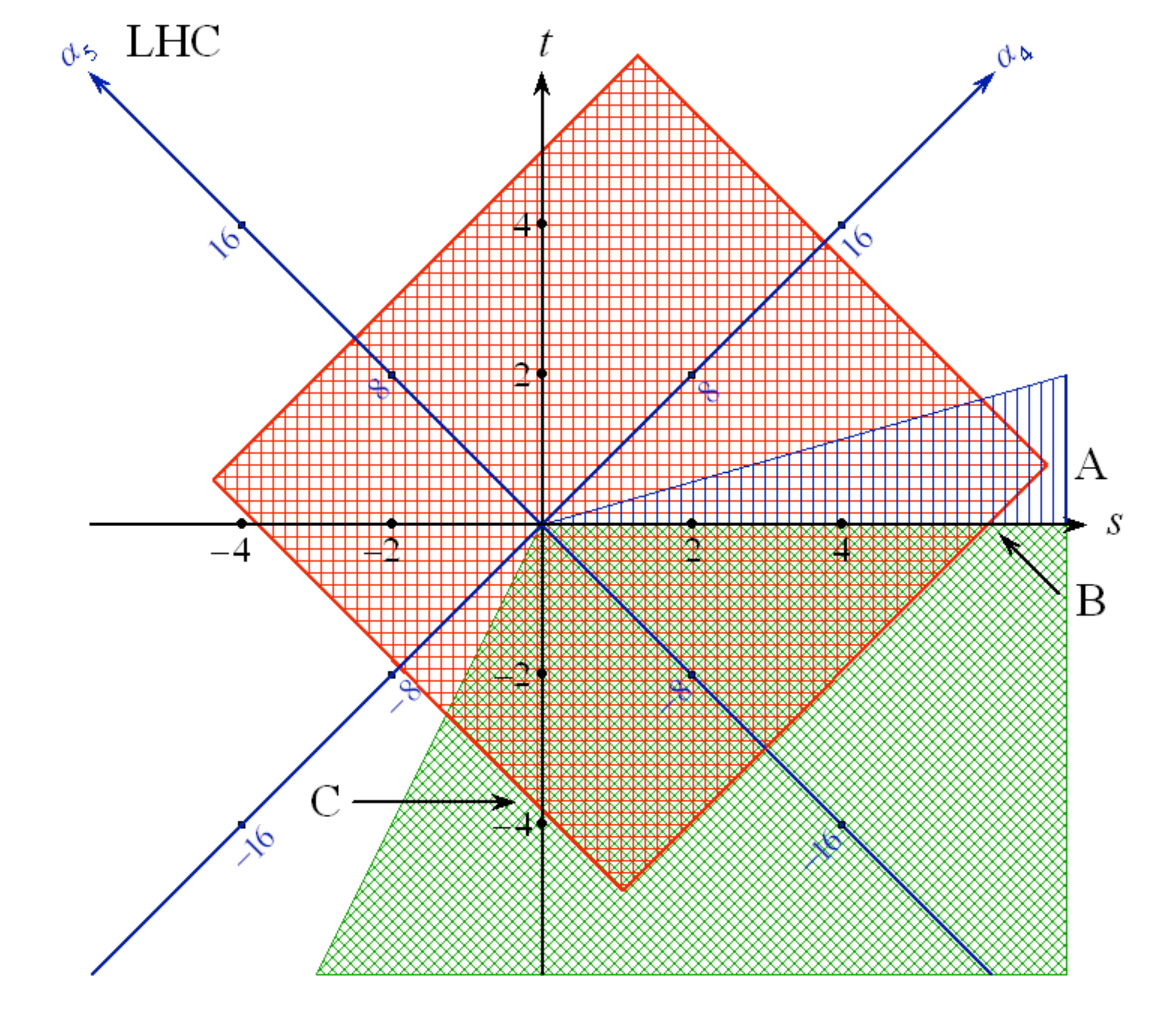} \quad
\caption{\it Comparison of  the estimated LHC sensitivity range~\cite{EGM} for
higher-order electroweak Lagrangian parameters with the 
the range of parameters where stable electroweak baryonic solitons
exist. The parameters $\alpha_{4}$, $\alpha_{5}$, $s$ and $t$ are in units of
$10^{-3}$. Point A has the maximal mass overall, Point B has the maximal 
mass allowed in the pure Skyrme limit: $t = 0$, and Point C has the
maximal mass allowed
in the limit where the Skyrme term vanishes: $s = 0$.}%
\label{figst_vs_alfas_LHC}%
\end{figure}

A study has been made of the prospective LHC sensitivity to the
higher-order electroweak Lagrangian terms. It was estimated that, in the absence of a
signal, the LHC could yield the following allowed ranges \cite{EGM}%
\begin{align}
-7.7\times10^{-3}  &  <\alpha_{4}<15\times10^{-3} \, ,\nonumber\\
-12\times10^{-3}  &  <\alpha_{5}<10\times10^{-3} \, . \label{possibleLHCbounds}%
\end{align}
These ranges are superimposed on the $(s,t)$ plane in Fig.~\ref{figst_vs_alfas_LHC}.
As in the previous examples, we list below the classical masses calculated for
three illustrative points that are also displayed in Fig.~\ref{figst_vs_alfas_LHC}:

\begin{itemize}
\item $\mathrm{A}: (s, t) = (6.75, 0.75)\,\times10^{-3}$ (maximal mass point):
$M_{\mathrm{A}}\simeq8.1$ TeV;

\item $\mathrm{B}: (s, t) = (6.0, \ \ 0.0\ )\,\times10^{-3}$ (Skyrme point): 
\qquad \ \ \,\,$M_{\mathrm{B}%
}\simeq7.9$ TeV;

\item $\mathrm{C}: (s, t) = (0.0, -3.85)\times10^{-3}$ (non-Skyrme point):
\quad $M_{\mathrm{C}}\simeq4.6\;$TeV.
\end{itemize}

The upper limit on the possible electroweak baryon mass in the event that
the LHC does not find non-zero higher-order effective Lagrangian parameters
is $\simeq 8$~TeV, very similar to the bound quoted in~\cite{EK}. This reflects
the fact that the point with maximal mass overall is quite close to the Skyrme
limit $t = 0$. In this case, unlike the previous examples,
the upper limit on the soliton mass is not relaxed
by allowing a non-Skyrme term with $t < 0$.

\section{Electroweak Baryons as Dark Matter?}
\label{dark}

If they exist, electroweak baryons should be present in the Universe today,
and could provide cold dark matter~\cite{Nussinov}. Their possible relic density today depends
on the presence and magnitude of a primordial electroweak baryon asymmetry.
If this was small, and the electroweak baryon density was completely
equilibriated in the early Universe, the present abundance of electroweak baryons
would be insufficient to provide the present density of cold dark matter. In this case,
one or more other sources of cold dark matter would be required, and
the constraints on dark matter scattering that we now discuss would be
irrelevant.

Assuming that they make up the bulk of the cold dark matter density,
it is clear that electroweak baryons should have no electric charge.
Moreover, it was pointed out in~\cite{Dine} that they could not be fermions, as
these would have an unacceptably large cross section for spin-independent scattering through 
magnetic moment couplings. Some scenarios for the effective low-energy
electroweak Lagrangian in which the lightest electroweak
baryon is a neutral boson have been enumerated in~\cite{Gillioz}, in an analysis based on
the Wess-Zumino-Witten term and the standard relation $Q_{\rm em} = I_3 + Y$. The neutral boson scenarios found {\it do not}
include SU(3) $\times$ SU(3) $\to$ SU(3) (which could yield a neutral fermion
if $N_{\rm c} = 3$ or a bosonic baryon if $N_{\rm c}$ is even), but {\it do} include
SU($N$) $\times$ SU($N$) $\to$ SU($N$) with $N > 3$ and $N_{\rm c}$ even,
{\it and} SO($N$) $\times$ SO($N$) $\to$ SO($N$).
However, SU($N$) $\to$ SO($N$) would yield a boson with charge $N_{\rm c}$, and the other
coset structures considered in~\cite{Gillioz} have trivial homotopy and hence
no electroweak baryons.

It should be noted that dark matter scattering is a non-trivial constraint, even
if the lightest electroweak baryon is a neutral boson. The scattering cross
section for this case was estimated in~\cite{CEO}, with results
indicating that models containing a pseudo-dilaton identified with the recently-discovered
boson $X$ with mass $\simeq 125$~GeV would be consistent with the (updated)
XENON100 limit~\cite{XENON100} if the pseudo-dilaton couplings were scaled by $\ge 1$~TeV~\footnote{This is
in the ball-park calculated in holographic dilaton models, and is reported to be consistent
with the available data on $X(125)$~\cite{MY}, once the contributions of the strongly-interacting 
electroweak sector the $Xgg$ and $X \gamma \gamma$ couplings are taken into account.}.
and the electroweak baryon weighed $\le 1$~TeV, as seen in Fig.~4 of~\cite{CEO}.
The XENON100 Collaboration does not report results for larger dark matter particle masses,
but naive extrapolation of the theoretical calculations and experimental sensitivity
would suggest some tension for masses above 1~TeV.

Our results suggest that scenarios which, according to the analysis of~\cite{Gillioz},
might yield either a charged bosonic baryon or a fermionic baryon {\it should not
necessarily be abandoned}. By the same token, one might not need to worry
about the prospective tension with XENON100. This is because the parameters of the fourth-order
effective Lagrangian might be in the range where no {\it stable} solitonic baryon
exists, i.e., $t > 0$ or $t < 0$ and $t/s > 2$. That said, it could be that stable
electroweak baryons nevertheless exist for such parameter choices, but cannot
be found using the solitonic Ansatz discussed here.

\section{Concluding Remarks}
\label{conx}

In this paper we have revisited the existence, stability and possible masses of classical solitonic
solutions to the effective low-energy Lagrangians of QCD and a possible strongly-interacting
electroweak sector, with particular attention to the presence and consequences of a
possible non-Skyrme quartic term as in (\ref{L4})~\cite{PR,FOT,LLVC,AAN}. We have revisited the stability
constraints in the $(s, t)$ plane, see Fig.~\ref{tsplane}, and given general results for the classical soliton mass,
see, e.g., Fig.~\ref{figlevels}.

In the QCD case, we have found that current phenomenological constraints on the
quartic chiral Lagrangian terms~\cite{EGPR,Pichfit} allow a range of classical baryon masses that is
somewhat broader than in the pure Skyrme case with $t = 0$. In the case of a possible strongly-interacting
electroweak sector, current data~\cite{EGM} allow a somewhat wider range of $t/s$ than in QCD, and the
expansion in the possible range of electroweak baryon masses is proportionally
larger than in the pure Skyrme case. On the other hand, the prospective LHC sensitivity~\cite{EGM} to quartic Lagrangian
parameters would not allow masses substantially larger than in the Skyrme limit.

It would be interesting to establish a `no-lose' theorem that the LHC will either
find non-zero higher-order Lagrangian parameters or discover electroweak
baryons. However, this would require greater LHC sensitivity than indicated
by the prospective range (\ref{possibleLHCbounds}). To our knowledge,
there is no accurate estimate of the range of electroweak baryon masses
that the LHC could detect, but it is surely less than the ranges quoted above
on the basis of a theoretical estimate of the possible LHC sensitivity to
higher-order Lagrangian terms.
On the other hand, the prospective LHC sensitivity should be revisited by
a full experimental simulation including the possibility of high-luminosity
LHC running.

As already commented, in order to be a valid dark matter candidate,
a stable electroweak solitonic baryon should be a neutral boson. Even
in this case, Fig.~4 of~\cite{CEO} indicates that such a model is quite
constrained, at least in models in which the recently-discovered new
boson is interpreted as a pseudo-dilaton. The estimate given in~\cite{CEO} of the
spin-independent cold dark matter scattering cross section rises with the
Skyrmion mass, so that experiments such as XENON100~\cite{XENON100} may be more sensitive
to models with heavier electroweak baryons. Full exploration of this issue lies
beyond the scope of this paper, but it is clear that dark matter scattering experiments
are potentially interesting probes of strongly-interacting electroweak models with
Skyrmion solutions.

\section*{Acknowledgements}

The work of J.E. was supported
partly by the London Centre for Terauniverse Studies (LCTS), using funding from the 
European Research Council via the Advanced Investigator Grant 267352. The work 
of M.P. was partly supported by the Polish NCN grant 2011/01/B/ST2/00492.


\begin{thebibliography}{99}

\bibitem{W}
E.~Witten,
\emph{Current Algebra, Baryons, and Quark Confinement},
  Nucl.\ Phys.\ B {\bf 223} (1983) 433.
  
\bibitem{ANW}
  G.~S.~Adkins, C.~R.~Nappi and E.~Witten,
	\emph{Static Properties of Nucleons in the Skyrme Model},
  Nucl.\ Phys.\ B {\bf 228} (1983) 552.

\bibitem{SW}
For a rigorous example, see
N.~Seiberg and E.~Witten,
\emph{Electric - magnetic duality, monopole condensation, and confinement in N=2 supersymmetric Yang-Mills theory},
  Nucl.\ Phys.\ B {\bf 426} (1994) 19
   [Erratum-ibid.\ B {\bf 430} (1994) 485]
  [hep-th/9407087] and
	\emph{Monopoles, duality and chiral symmetry breaking in N=2 supersymmetric QCD},
  Nucl.\ Phys.\ B {\bf 431} (1994) 484
  [hep-th/9408099].

\bibitem{Skyrme}
T.~H.~R.~Skyrme,
\emph{A Nonlinear field theory},
  Proc.\ Roy.\ Soc.\ Lond.\ A {\bf 260} (1961) 127 and
\emph{A Unified Field Theory of Mesons and Baryons},
  Nucl.\ Phys.\  {\bf 31} (1962) 556.
 
\bibitem{SkyrmeOK}
G.~S.~Adkins and C.~R.~Nappi,
\emph{The Skyrme Model with Pion Masses},
  Nucl.\ Phys.\ B {\bf 233} (1984) 109 and
\emph{Model Independent Relations For Baryons As Solitons In Mesonic Theories},
  Nucl.\ Phys.\ B {\bf 249} (1985) 507;\\
 M.~Karliner and M.~P.~Mattis,
\emph{Hadron Dynamics In The Three Flavor Skyrme Model}
  Phys.\ Rev.\ Lett.\  {\bf 56} (1986) 428 and
\emph{The Baryon Spectrum of the Skyrme Model},
  Phys.\ Rev.\ D {\bf 31} (1985) 2833 and
\emph{$\pi$N, KN and anti-KN Scattering: Skyrme Model Versus Experiment},
  Phys.\ Rev.\ D {\bf 34} (1986) 1991.
  
\bibitem{Meissner:1987ge}
  U.~G.~Meissner,
\emph{Low-Energy Hadron Physics from Effective Chiral Lagrangians with Vector Mesons},
  Phys.\ Rept.\  {\bf 161} (1988) 213.
  
\bibitem{Weigel}
H.~Weigel,
\emph{Chiral Soliton Models for Baryons},
  Lect.\ Notes Phys.\  {\bf 743} (2008) 1.
  
\bibitem{TC}
S.~Weinberg,
\emph{Implications Of Dynamical Symmetry Breaking},
  Phys.\ Rev.\  D {\bf 13} (1976) 974;\\
  L.~Susskind,
\emph{Dynamics Of Spontaneous Symmetry Breaking In The Weinberg-Salam Theory},
  Phys.\ Rev.\  D {\bf 20} (1979) 2619.

\bibitem{X}
G.~Aad {\it et al.}  [ATLAS Collaboration],
\emph{Observation of a new particle in the search for the Standard Model Higgs boson with the ATLAS detector at the LHC},
  Phys.\ Lett.\ B {\bf 716} (2012) 1
  [arXiv:1207.7214 [hep-ex]];\\
  S.~Chatrchyan {\it et al.}  [CMS Collaboration],
\emph{Observation of a new boson at a mass of 125 GeV with the CMS experiment at the LHC},
  Phys.\ Lett.\ B {\bf 716} (2012) 30
  [arXiv:1207.7235 [hep-ex]].

\bibitem{PNGB}
G.~F.~Giudice, C.~Grojean, A.~Pomarol and R.~Rattazzi,
\emph{The Strongly-Interacting Light Higgs},
  JHEP {\bf 0706} (2007) 045
  [hep-ph/0703164];\\
R.~Contino, C.~Grojean, M.~Moretti, F.~Piccinini and R.~Rattazzi,
 \emph{Strong Double Higgs Production at the LHC},
  JHEP {\bf 1005} (2010) 089
  [arXiv:1002.1011 [hep-ph]];\\
  R.~Contino,
\emph{Tasi 2009 lectures: The Higgs as a Composite Nambu-Goldstone Boson},
arXiv:1005.4269 [hep-ph];\\
 R.~Grober and M.~Muhlleitner,
  \emph{Composite Higgs Boson Pair Production at the LHC},
  JHEP {\bf 1106} (2011) 020
  [arXiv:1012.1562 [hep-ph]].

\bibitem{Dilaton}
B.~Holdom,
\emph{Techniodor},
  Phys.\ Lett.\  B {\bf 150} (1985) 301;\\
K.~Yamawaki, M.~Bando and K.~i.~Matumoto,
\emph{Scale Invariant Technicolor Model And A Technidilaton},
  Phys.\ Rev.\ Lett.\  {\bf 56} (1986) 1335;\\
  T.~W.~Appelquist, D.~Karabali and L.~C.~R.~Wijewardhana,
\emph{Chiral Hierarchies and the Flavor Changing Neutral Current Problem in
 Technicolor},
  Phys.\ Rev.\ Lett.\  {\bf 57} (1986) 957;\\
  T.~Akiba and T.~Yanagida,
\emph{Hierarchic Chiral Condensate},
  Phys.\ Lett.\ B {\bf 169} (1986) 432;\\
 M.~Bando, K.~-i.~Matumoto and K.~Yamawaki,
\emph{Technidilaton},
Phys.\ Lett.\ B {\bf 178}, 308 (1986);\\
 W.~D.~Goldberger, B.~Grinstein and W.~Skiba,
\emph{Distinguishing the Higgs boson from the dilaton at the Large Hadron
 Collider},
  Phys.\ Rev.\ Lett.\  {\bf 100} (2008) 111802
  [arXiv:0708.1463 [hep-ph]];\\
J.~Fan, W.~D.~Goldberger, A.~Ross and W.~Skiba,
  \emph{Standard Model couplings and collider signatures of a light scalar},
  Phys.\ Rev.\  D {\bf 79} (2009) 035017
  [arXiv:0803.2040 [hep-ph]];\\
  L. Vecchi, 
  \emph{Phenomenology of a light scalar: the dilaton},
    Phys.~Rev.~D {\bf 82} (2010) 076009;\\
  B.~Grinstein and P.~Uttayarat,
\emph{A Very Light Dilaton},
  JHEP {\bf 1107} (2011) 038
  [arXiv:1105.2370 [hep-ph]];\\
  S.~Matsuzaki and K.~Yamawaki,
\emph{Techni-dilaton signatures at LHC},
Prog.\  Theor.\  Phys.\  {\bf 127}, 209 (2012)
[arXiv:1109.5448 [hep-ph]];\\
V.~Barger, M.~Ishida and W.~Keung,
  \emph{Dilaton at the LHC},
  arXiv:1111.2580 [hep-ph];\\
B.~Coleppa, T.~Gregoire and H.~E.~Logan,
  \emph{Dilaton constraints and LHC prospects},
  arXiv::1111.3276 [hep-ph];\\
C.~Coriano, L.~Delle Rose, A.~Quintavalle and M.~Serino,
 \emph{Effective Dilaton Interactions from the Anomalous Breaking of 
 Scale Invariance of the Standard Model},
  arXiv:1206.0590 [hep-ph];\\
  S.~Matsuzaki and K.~Yamawaki,
\emph{Discovering 125 GeV techni-dilaton at LHC},
  Phys.\ Rev.\ D {\bf 86} (2012) 035025
  [arXiv:1206.6703 [hep-ph]];\\
  \emph{C.~Coriano, L.~Delle Rose, C.~Marzo and M.~Serino},
  \emph{Higher Order Dilaton Interactions in the Nearly Conformal Limit of the Standard Model},
  arXiv:1207.2930 [hep-ph];\\
  S.~Matsuzaki and K.~Yamawaki,
  \emph{Is 125 GeV techni-dilaton found at LHC?},
  arXiv:1207.5911 [hep-ph];\\
  D.~Elander and M.~Piai,
 \emph{The decay constant of the holographic techni-dilaton and the 125 GeV boson},
  arXiv:1208.0546 [hep-ph];\\
  L.~Delle Rose and M.~Serino,
  \emph{Dilaton Interactions in QCD and in the Electroweak Sector of the Standard Model},
  arXiv:1208.6432 [hep-ph];\\
  \emph{S.~Matsuzaki and K.~Yamawaki},
  \emph{Holographic techni-dilaton at 125 GeV},
  arXiv:1209.2017 [hep-ph];\\
  Z.~Chacko and R.~K.~Mishra,
\emph{Effective Theory of a Light Dilaton},
  arXiv:1209.3022 [hep-ph];\\
  Z.~Chacko, R.~Franceschini and R.~K.~Mishra,
  \emph{Resonance at 125 GeV: Higgs or Dilaton/Radion?},
  arXiv:1209.3259 [hep-ph];\\
  \emph{B.~Bellazzini, C.~Csaki, J.~Hubisz, J.~Serra and J.~Terning},
\emph{A Higgslike Dilaton},
  arXiv:1209.3299 [hep-ph].
  
\bibitem{EK}
J.~Ellis and M.~Karliner,
\emph{Indications on the Mass of the Lightest Electroweak Baryon},
  Phys.\ Lett.\ B {\bf 713} (2012) 233
  [arXiv:1204.6642 [hep-ph]].

\bibitem{PR}
D.~E.~L.~Pottinger and E.~Rathske,
\emph{Metastability Of Solitons In A Generalized Skyrme Model},
Phys.\ Rev.\ D \textbf{33} (1986) 2448;\\
E.~Rathske,
\emph{Research on phenomenological models with chiral symmetry},
BONN-IR-91-70.

\bibitem{FOT}
K.~Fujii, S.~Otsuki and F.~Toyoda,
\emph{Solitons With The Hopf Index Versus Skyrmions In Su(2) Nonlinear Sigma Model},
Prog.\ Theor.\ Phys.\ \textbf{73} (1985) 1287.

\bibitem{LLVC}
M.~Lacombe, B.~Loiseau, R.~Vinh Mau and W.~N.~Cottingham,
\emph{Nuclear Binding And The Skyrme Model},
Phys.\ Lett.\ B \textbf{161} (1985) 31.

\bibitem{DGH}
J.~F.~Donoghue, E.~Golowich and B.~R.~Holstein,
\emph{Predicting The Proton Mass from Pi Pi Scattering Data},
Phys.\ Rev.\ Lett.\ \textbf{53} (1984) 747.

\bibitem{AAN}
A.~A.~Andrianov, V.~A.~Andrianov and V.~Y.~.Novozhilov,
\emph{Comment On 'Predicting The Proton Mass from Pi Pi Scattering Data'},
Phys.\ Rev.\ Lett.\ \textbf{56} (1986) 1882.

\bibitem{Aitchison:1986aq}
  I.~Aitchison, C.~Fraser, E.~Tudor and J.~Zuk,
\emph{Failure Of The Derivative Expansion For Studying Stability Of The Baryon As A Chiral Soliton},
  Phys.\ Lett.\ B {\bf 165} (1985) 162.
  
\bibitem{Adkins:1983nw} 
  G.~S.~Adkins and C.~R.~Nappi,
\emph{Stabilization of Chiral Solitons via Vector Mesons},
  Phys.\ Lett.\ B {\bf 137}, 251 (1984).
  
\bibitem{Mashaal:1985rg}
  M.~Mashaal, T.~N.~Pham and T.~N.~Truong,
\emph{Can Skyrmion Be A Good Description Of Nucleon?},
  Phys.\ Rev.\ Lett.\  {\bf 56} (1986) 436.
  
\bibitem{Moussallam:1992ia}
  B.~Moussallam,
\emph{Casimir energy in the Skyrme model},
  Conf.\ Proc.\ C {\bf 9209271} (1992) 269
  [hep-ph/9211229].
  
\bibitem{Holzwarth:1995bv}
  G.~Holzwarth and H.~Walliser,
\emph{Quantum corrections to the skyrmion mass},
  Nucl.\ Phys.\ A {\bf 587} (1995) 721.

\bibitem{EGPR}
G.~Ecker, J.~Gasser, A.~Pich and E.~de Rafael,
\emph{The Role Of Resonances In Chiral Perturbation Theory},
  Nucl.\ Phys.\  B {\bf 321} (1989) 311.

\bibitem{Pichfit}
A.~Pich,
\emph{Low-Energy Constants from Resonance Chiral Theory},
  PoS Confinement {\bf 8} (2008) 026
  [arXiv:0812.2631 [hep-ph]].

\bibitem{EGM}
O.~J.~P.~Eboli, M.~C.~Gonzalez-Garcia and J.~K.~Mizukoshi,
\emph{$p p \rightarrow j j e^{\pm} \mu^{\pm} \nu \nu$ and  $j j e^{\pm} \mu^{\mp} \nu \nu$ at 
${\cal O}(\alpha_{\rm em}^6)$
 and
 ${\cal O}(\alpha_{\rm em}^4 \alpha_{\rm s}^2)$ for the study of the quartic electroweak
  gauge boson vertex at LHC},
  Phys.\ Rev.\  D {\bf 74} (2006) 073005
  [arXiv:hep-ph/0606118].

 \bibitem{DGPR}
J.~Distler, B.~Grinstein, R.~A.~Porto and I.~Z.~Rothstein,
\emph{Falsifying Models of New Physics Via WW Scattering},
  Phys.\ Rev.\ Lett.\  {\bf 98} (2007) 041601
  [arXiv:hep-ph/0604255].

\bibitem{Nussinov}
S.~Nussinov,
\emph{Technocosmology: Could A Technibaryon Excess Provide A 'natural' Missing Mass Candidate?},
  Phys.\ Lett.\ B {\bf 165} (1985) 55.
  
\bibitem{CEO}
B.~A.~Campbell, J.~Ellis and K.~A.~Olive,
\emph{Phenomenology and Cosmology of an Electroweak Pseudo-Dilaton and Electroweak Baryons},
  JHEP {\bf 1203} (2012) 026
  [arXiv:1111.4495 [hep-ph]].
  
\bibitem{Pham:1985cr}
  T.~N.~Pham and T.~N.~Truong,
\emph{Evaluation Of The Derivative Quartic Terms Of The Meson Chiral Lagrangian From Forward Dispersion Relation},
  Phys.\ Rev.\ D {\bf 31} (1985) 3027.

  
\bibitem{XENON100}
E.~Aprile {\it et al.}  [XENON100 Collaboration],
\emph{Dark Matter Results from 225 Live Days of XENON100 Data},
  arXiv:1207.5988 [astro-ph.CO].
  
\bibitem{Dine}
J.~Bagnasco, M.~Dine and S.~D.~Thomas,
\emph{Detecting technibaryon dark matter},
  Phys.\ Lett.\  B {\bf 320} (1994) 99
  [arXiv:hep-ph/9310290].
  
\bibitem{Gillioz}
M.~Gillioz,
\emph{Dangerous Skyrmions in Little Higgs Models},
  JHEP {\bf 1202} (2012) 121
  [arXiv:1111.2047 [hep-ph]].

\bibitem{MY}
See the last paper by S.~Matsuzaki and K.~Yamawaki in~\cite{Dilaton}.

\end{thebibliography}
\end{document}